# Building 3D superconductor-based Josephson junctions using a via transfer approach

Cequn Li[1,#], Le Yi[1,2], Kalana D. Halanayake[3], Jessica L. Thompson[3], Yingdong Guan[1,4], Kenji Watanabe[5], Takashi Taniguchi[6], Zhiqiang Mao[1,4], Danielle Reifsnyder Hickey[2,3,7], Morteza Kayyalha[2,8], Jun Zhu[1,2*]

1. Department of Physics, The Pennsylvania State University, University Park, PA, USA
2. Materials Research Institute, The Pennsylvania State University, University Park, PA 16802, USA
3. Department of Chemistry, The Pennsylvania State University, University Park, PA 16802, USA
4. 2-Dimensional Crystal Consortium, The Pennsylvania State University, University Park, PA 16802, USA
5. Research Center for Electronic and Optical Materials, National Institute for Materials Science, 1-1 Namiki, Tsukuba 305-0044, Japan
6. Research Center for Materials Nanoarchitectonics, National Institute for Materials Science, 1-1 Namiki, Tsukuba 305-0044, Japan
7. Department of Materials Science and Engineering, The Pennsylvania State University, University Park, PA 16802, USA
8. Department of Electrical Engineering, The Pennsylvania State University, University Park, Pennsylvania 16802, USA

\# Present address: Laboratory of Atomic and Solid State Physics, Cornell University, Ithaca, NY 14853, USA

\* Correspondence to: jxz26@psu.edu (J. Zhu)

**Abstract:** The coupling of superconductivity to unconventional materials may lead to novel quantum states and potential applications. Controlling the quality of the superconductor-normal metal interface is of crucial importance to the understanding and engineering of the superconducting proximity effect. In many cases, conventional lithography-based deposition methods introduce undesirable effects. Using the concept of via contact and dry transfer, we have constructed smooth, van der Waals-like contact between 3D superconducting NbN/Pd and graphene with low contact resistance of approximately $130\ \Omega \cdot \mu m$. Gate-tunable supercurrent, Fraunhofer pattern, and Andreev reflections are observed, the properties of **which can be understood using an induced superconducting gap $\Delta'$ in this planar contact geometry. We discuss potential mechanisms impacting the magnitude of $\Delta'$ and suggest ways of further increasing the proximity coupling. This gentle, lithography-free contacting method can be applied to air- and damage-sensitive surfaces to engineer novel superconducting heterostructures.**

**Main text:** The superconducting proximity effect in two-dimensional (2D) materials and surfaces offers a promising avenue to explore emergent quantum phenomena such as topological superconductivity [1-14], 0-π Josephson junctions [15, 16], and their potential applications in quantum computing [1]. Research on graphene Josephson junctions (JJ) has been particularly fruitful [4, 12, 17-34], with the realization of ballistic JJs [19, 20, 24] and evidence for superconducting correlation and crossed Andreev reflections in the quantum Hall regime [4, 5, 12]. Because graphene is gapless, robust, and chemically inert, it forms an ohmic contact with many conventional 3D superconductors (SC) deposited either on top [17-19, 22, 26, 34] or on the edge [4, 5, 12, 20, 24, 25, 28, 30, 32] using conventional evaporation or sputtering techniques. This is not the case for more sensitive surfaces such as a transition metal dichalcogenide (TMD) semiconductor [35] or a topological insulator (TI). The direct sputtering of Nb on $BiSbTe_2Se$, for example, requires in-situ Ar-plasma surface cleaning and is known to cause deformation and form amorphous layers at the SC-TI interface [11]. These complications necessitate a gentler contact approach, such as the van der Waals transfer [35]. Indeed, highly transparent interfaces have been achieved in a NbSe$_2$/bilayer graphene SC-normal metal (SN) junction [36], and lateral and vertical JJs using NbSe$_2$ have also been reported [21, 29, 31, 33]. However, this approach is limited to primarily NbSe$_2$, which is a van der Waals SC.

Previous studies have shown that it is possible to form a van der Waals-like interface between a 3D metal and a 2D metal or semiconductor by first depositing the metal into pits etched into an h-BN flake, then lifting and transferring the metal-embedded h-BN flake onto the target material [37-39]. This via transfer approach can achieve a contact resistance to graphene and TMD materials (such as NbSe$_2$, MoTe$_2$, and WSe$_2$) comparable to standard deposition techniques [37, 38]. More importantly, it is polymer-free, avoids surface damage, and can be carried out in a glovebox, making it highly suitable for air-sensitive and damage-prone surfaces [37, 38]. The realization of superconducting via contacts has clear advantages. Indeed, previous efforts have tried to make superconducting via contacts to ZrTe$_5$ [40] and MoS$_2$ [41], but a Josephson current has not been observed, likely due to low interface transparency [40, 41].

In this work, we report on the successful fabrication and studies of NbN/Pd-graphene lateral JJs using the via contact approach. NbN/Pd via contacts are made to monolayer or bilayer graphene and the transfer results in a smooth and tightly conforming van der Waals-like interface, with a unit-length contact resistance of $\sim 130\ \Omega \cdot \mu m$, comparable to the conventional deposition techniques [12, 19, 22, 24, 37]. Gate-tuned critical supercurrent on the order of 0.1 µA and Andreev reflections are observed in our devices and their temperature and magnetic field dependences follow the standard descriptions of a Josephson junction very well. Our measurements suggest that a proximity-induced gap $\Delta'$ of several tens of µeV develops in the graphene area beneath the NbN/Pd contact, the magnitude of which is influenced by the strong disorder in the NbN film. Our work establishes via contacts as a viable approach to constructing superconducting devices on 2D materials and sensitive surfaces, and points to future improvement strategies.

We construct the via contact stack following Ref. [37] and illustrate this process flow in Fig. 1a. We first etch pits in an h-BN flake exfoliated on a SiO$_2$/Si substrate and then sputter NbN/Pd film (45/4 nm) into the pits. Here, the insertion of a thin Pd layer prevents the NbN film from sticking to the SiO$_2$ substrate [37]. Pd is used as a buffer layer in the formation of a highly transparent NbN/Pd-graphene interface [19] and makes good electrical contact to graphene [23]. We assemble the stack using a dry transfer technique [13, 36] by picking up in sequence the NbN/Pd-embedded h-BN flake, a monolayer graphene (MLG) or bilayer graphene (BLG) flake, an h-BN dielectric flake, and a graphite flake serving as the backgate. Before assembly, the NbN/Pd film is protected from oxidation by the surrounding h-BN structure. The assembly is performed inside an Ar-filled glovebox. Standard lithography is used to complete the device. The complete fabrication details are given in Supplementary Section 1. Figure 1b shows an optical image of an exemplary device, J008. Each NbN/Pd strip measures 0.3 μm by 7 μm, separated by a BLG channel that measures $L = 0.69$ μm in length and $W = 2.9$ μm in width. In J008, the NbN/Pd contacts exhibit a superconducting transition temperature $T_c$ of ~ 7 K. A comprehensive study of our sputtered NbN/Pd films is given in Supplementary Section 2. In general, our films exhibit $T_c$ in the range of 6–11 K, and an upper critical field of $H_{c2} > 14$ T at 2 K.

We first examine the via contact-graphene interface using scanning and transmission electron microscopy (STEM), energy-dispersive X-ray spectroscopy (EDX), electron energy-loss spectroscopy (EELS), and atomic force microscopy (AFM). Figures 1d and e show the highlights of our findings. Figure 1d shows an annular dark-field (ADF-) STEM cross-sectional image of one NbN/Pd/BLG junction (with a magnified image of one side), which was lifted from a device similar to J008 using a focused ion beam (see Supplementary Section 3 for details of the STEM sample preparation and additional measurements). The etched h-BN pits have sloped side walls and are completely filled by the sputtered film without any void along the entire length of the device (see Supplementary Fig. S3a for a low-magnification image of the entire device). The overlaid EDX/EELS maps shown in Fig. 1e confirm the sharp and tightly conforming Pd/BLG interface in our device, similar to previous via contacts made using Au, Pt, or Pd [37, 39]. AFM measurements given in Supplementary Fig. S3 show that the Pd side of a sputtered NbN/Pd contact has a small root-mean-square (RMS) roughness of 1.8 Å, similar to the $SiO_2$ substrate it is deposited on. The smoothness of the film facilitates the formation of a van der Waals-like interface in Fig. 1d and reduces interface scattering; this helps improve the proximity coupling [36, 40]. We have also transferred our NbN/Pd via contacts to $Bi_2Se_3$, an air-sensitive topological insulator, and obtained smooth and bubble-free stacks (See Supplementary Fig. S11).

Figure 2a plots the junction resistance $R_J$ at $T = 3$ K (black trace) in device J008 as a function of the backgate voltage $V_g$ measured using the JJ configuration shown in Fig. 1c. Also shown for comparison is a four-terminal $R_{xx}(V_g)$ obtained in the pristine BLG region (red trace). The resistance peak at $V_g = -0.15$ V corresponds to the charge neutrality point (CNP) of the BLG channel between the NbN/Pd contacts since a peak at the same $V_g$ occurs in pristine BLG (CNP 1).

We attribute the resistance peak at $V_g = 0.26$ V to the CNP of the BLG regions underneath the NbN/Pd contacts (CNP 2). Using the quantum Hall effect (Fig. 2b), we determine that the NbN/Pd contact dopes the BLG underneath with an average hole level of $2.56 \times 10^{11}$ cm$^{-2}$. A similar hole doping level of $1.5 \times 10^{11}$ cm$^{-2}$ was found for device J004. It follows from the black trace in Fig. 2a that the carrier density profile evolves from p'-p-p' to p'-n-p', and eventually to n'-n-n' as $V_g$ increases, as illustrated in the insets. This picture is further supported by the Landau fan diagram in Supplementary Fig. S4a. In Fig. 2b, $R_J$ of the quantum Hall resistance plateau is given by

$$R_J = \frac{h}{e^2} \times \frac{1}{|\nu|} + 2R_C, \qquad (\text{Eq. 1})$$

where $h/e^2$ is the von Klitzing constant, $\nu$ is the filling factor, and $R_C$ the contact resistance for one NbN/Pd contact [42]. Fits to Eq. 1 are shown in Fig. 2c, from which we determine $R_C = 45 \pm 4$ Ω on the hole side and extract a unit-length contact resistance of $\rho_C = 126 \pm 11$ Ω·µm. This value is comparable to other superconducting top or edge contacts on graphene [12, 19, 22, 24], and to the Au via contacts [37]. The small $R_C$ suggests a well-coupled Pd/BLG interface [36], consistent with our STEM findings in Fig. 1. $R_C$ is considerably larger when the NbN/Pd/BLG contact area is electron-doped, likely due to the inhomogeneous hole doping of the NbN/Pd film, and the formation of many local p-n junctions [34, 43].

A supercurrent develops as the temperature drops to below 0.5 K or so. Figure 3a shows a false-colored map of the differential resistance $dV/dI$ in device J008 as a function of the dc current bias $I_{dc}$ and the backgate voltage $V_g$. The critical supercurrent $I_c$ – defined as the current corresponding to a small threshold voltage $V_c = 0.5$ µV (Ref. [33]) – grows steadily with increasing hole doping and reaches approximately 110 nA at $V_g = -4$ V or a hole density $p = 2.4 \times 10^{12}$ cm$^{-2}$; This corresponds to a critical current density $J_c$ of 40 nA · µm$^{-1}$. $I_c$ vanishes when the NbN/Pd/BLG contact region reaches charge neutral and recovers partially when the device enters the n'-n-n' regime, albeit with a much smaller amplitude of ~ 20 nA. Figure 3b plots the $I_c R_N$ product, a common criterion to assess the efficiency of the superconducting proximity effect [27, 44], where the normal state resistance $R_N$ is obtained from the d$I$/d$V(V_{dc})$ plot at large bias (See Supplementary Fig. S5). $I_c R_N$ saturates to approximately 16 µV / 10 µV in the hole/electron regime, while the superconducting gap of the NbN/Pd film $\Delta_{NbN}$ in our devices is approximately 0.9 meV, obtained from the differential conductance measurements shown in Supplementary Fig. S5.

The supercurrent is nearly uniform across the width of the BLG channel, as indicated by the periodic modulation of $I_c$ in a magnetic field $B$ (Fig. 3c). The standard Fraunhofer pattern given by

$$I_c(B) = I_c(0) \frac{\Phi_0}{\pi B A_{\text{eff}}} \sin\left(\frac{\pi B A_{\text{eff}}}{\Phi_0}\right), \tag{Eq. 2}$$

where $\Phi_0 = h/2e$ is the magnetic flux quantum and $A_{\text{eff}}$ is the effective junction area [44], is plotted as the blue solid line in Fig. 3c and provides an excellent description of our data. The extracted $A_{\text{eff}} = 2.9$ µm$^2$ is $V_g$-independent and reaches excellent agreement with a calculated effective junction area $A_{\text{eff}} = W \times (L + 2L_{\text{S,half}}) = 2.87$ µm$^2$, where $L_{\text{S,half}} = 0.15$ µm is the half-length of the NbN/Pd electrodes due to the focusing of the magnetic flux expelled from the SC [4], $L = 0.69$ µm and $W = 2.9$ µm are the length and width of the BLG channel, respectively.

The fast decay of the side lobes may suggest a slight non-uniformity in the current distribution, due to inhomogeneous doping of the contact area.

Consistent results are obtained in another JJ device J004, where we observed $I_c$ of tens of nA, $I_c R_N$ of 10–25 µV, and a Fraunhofer pattern similar to Fig. 3c (see Supplementary Fig. S7). These findings show that the via method works reliably for NbN/Pd contacts and a uniform superconducting proximity coupling can be achieved using the transfer technique.

To further understand the behavior of the JJ and the impact of the via contact, we measured the current-voltage ($I$-$V$) characteristics of the junction at different temperatures and obtained the $T$-dependent $I_c$ at representative backgate voltages. The $I$-$V$ and $I_c(T)$ obtained from J008 are plotted in Figs. 3d and e respectively, while data from J004 are given in Supplementary Fig. S7. We estimated the mean free path $L_{\text{mfp}}$ of the BLG to be 100–200 nm for the holes (see Supplementary Section 4). $L_{\text{mfp}}$ is short compared to pristine BLG, likely due to the diffusion of chemically doped carriers from the contact area. Thus, our devices operate in the diffusive regime with the Thouless energy given by $E_{\text{Th}} = \hbar D/L^2$, and the superconducting coherence length $\xi$ given by $\xi = \sqrt{\hbar D/\Delta_{\text{NbN}}}$. Here, $D = v_F L_{\text{mfp}}/2$ is the 2D diffusion constant and $v_F = 1 \times 10^6$ m/s is the Fermi velocity in graphene [27]. $E_{\text{Th}}$ is approximately 200 µeV and $L/\xi$ ranges from 1.5 to 2.3 in the hole regime, placing our JJs in between the short ($L/\xi < 1$) and long junction ($L/\xi > 10$) limits [18, 29, 31, 45].

Empirically, we find our $I_c(T)$ data can be described very well by the long diffusive model given by

$$eI_cR_N = \alpha_1 E_{Th}\left[1 - b\exp\left(\frac{-\alpha_2 E_{Th}}{3.2 k_B T}\right)\right], \qquad (Eq.\,3)$$

where $\alpha_1$, $\alpha_2$, and $b$ are fitting parameters [18, 29, 31, 45]. Fits to Eq. 3 (blue solid lines in Fig. 3e) yield $\alpha_1$, which represents the ratio $eI_cR_N/E_{Th}$ in the $T=0$ limit, of approximately 0.1 for holes (plotted on the right axis in Fig. 3b). They are much lower than the value of 10.82 expected for long diffusive JJs [45], but are in line with the values of $\alpha_{1,2}$ obtained in the literature [18, 22, 23, 29-31]. Supplementary Fig. S8 summarizes results of $eI_cR_N/E_{Th}$ obtained in long diffusive graphene JJs, where $\alpha_1$ systematically increases with increasing $L/\xi$. Deviations from theory are partly due to devices not reaching the true long junction limit [18, 29]. However, the strength of the SC proximity effect must also play a role [22, 23, 46].

To further understand what controls the strength of the proximity coupling, we measured the differential resistance of a single SN junction $dV/dI$ vs $I_{dc}$ using the SN configuration illustrated in Fig. 1c. A normal electron incident at a fully transparent S-N interface (barrier strength $Z=0$) undergoes a perfect Andreev reflection (AR), which leads to a two-fold conductance enhancement [47]. As $Z$ increases, the AR process is suppressed and replaced by electron tunneling behavior [9, 34, 36]. Our earlier work showed that a van der Waals NbSe$_2$/BLG interface supports nearly perfect AR with an enhancement factor of 1.8 [36]. Figure 4a shows the measured $dV/dI(I_{dc})$ in J008 at $V_g = -4$ V and selected temperatures (Large-range $I_{dc}$ sweeps and measurements at other $V_g$'s are given in Supplementary Fig. S9). The AR process manifests in the resistance reduction at small current bias $I_{dc} < I_{SN}$, where $I_{SN}$ marks the location of the $dV/dI$ peaks in Fig. 4a. Figure 4b plots the temperature dependence of $I_{SN}$ and $I_{JJ}$, which marks the location of the $dV/dI$ peaks in Fig. 3d. $I_{JJ}$ signals the onset of a supercurrent in the JJ; $I_{SN}$ tracks $I_{JJ}$ very well, indicating a common origin.

We attribute $I_{SN}$ to the development of a proximity-induced SC gap $\Delta'$ in the NbN/Pd/BLG contact region, following previous examples where the SC contact and the normal metal have a significant planar overlap area, e.g. NbSe$_2$ on graphene [36] or WTe$_2$ [7], Nb on Cd$_3$As$_2$ [9] or HgTe [3], or Al on InAs [6]. Near $V_{dc} = 0$, the behavior of a gapped BLG ($\Delta'$) -normal BLG-gapped BLG ($\Delta'$) dominates, as illustrated in Fig. 4e. Using the Blonder-Tinkham-Klapwijk (BTK) analysis [7, 9, 13, 34, 47], we extracted $\Delta'$ and the effective barrier strength $Z_{eff}$ at a number of $V_g$'s (See details in Supplementary Section 9). An exemplary fit is shown in Fig. 4c while the fitting results are given in Fig. 4d. $\Delta'$ is approximately 10 μeV in J008, and nearly $V_g$ independent on the hole side (See Fig. 4d). $eI_cR_N$ is approximately 1–2 $\Delta'$, indicating that the magnitude of the supercurrent is limited by the induced gap in the contact region, rather than the parent gap of NbN. This also explains our observation of $eI_cR_N \ll E_{Th}$ in the diffusive junction analysis, since a long dwell time of order $\tau_I = \hbar/\Delta'$ at the SC/graphene interface dominates the channel dwell time $\tau_d = \hbar/E_{Th}$ [22, 23, 46]. The BTK analysis also produced $Z_{eff}$ in the range of 0.3–0.5 and an interface transparency $\mathcal{T} = 1/(1 + Z_{eff}^2) \in [0.8, 0.92]$ [12, 34], indicating a high transparency between the normal BLG and gapped BLG. This is encouraging for experiments seeking to construct Josephson junctions within the same material [14, 48, 49].

Effectively, the JJs studied here can be viewed as two vertical SN junctions and a lateral JJ connected in series, as illustrated in Fig. 4e. The vertical junction is between the NbN/Pd and the BLG underneath, and the lateral JJ consists of gapped BLG-normal BLG-gapped BLG. In this view, the JJ is in the short diffusive regime of $L/\xi' < 1$, where $\xi' = \sqrt{\hbar D/\Delta'}$. Indeed, an alternative analysis of $I_c(T)$ using a short diffusive model [26, 50] (See details in Supplementary

Section 6) also provides an excellent description of our data. An exemplary fit is shown as an orange solid line in Fig. 3e. Similar to the case of Nb/HgTe in Ref. [3], $\Delta'$ extracted from this analysis is only a fraction of $\Delta_{NbN}$. It is in the range of 20–30 µeV, which is comparable to $eI_cR_N$ and approximately 2–3 larger than $\Delta'$ obtained from the BTK analysis of the SN junction. These consistent estimates further confirm the validity of the effective gap model.

Alternatively, we have considered the scenario where the graphene channel approaches ballistic transport in the hole regime and fit $I_c(T)$ to a short ballistic model [21]. In the temperature range we studied, the short ballistic model can also describe our data very well. The green solid line in Fig. 3e shows a representative fit at $V_g = -4$ V while other details are given in Supplementary Section 6. We obtained $\Delta'$ in the range of 30–50 µeV, or ~ 50% higher than $\Delta'$ obtained using the short diffusive model. Both models capture the $I_c(T)$ behavior of the gapped BLG-normal BLG-gapped BLG junction. In addition, the short ballistic fits yielded an interface transparency $\mathcal{T} > 0.9$, in agreement with $\mathcal{T} \sim 0.9$ obtained from the BTK analysis in Fig. 4d. These consistent results obtained from different analyses further establish the robustness of our conclusion.

Our experiment highlights the importance of increasing the induced gap $\Delta'$ in planar contact geometries. In the literature, $\Delta'$ reaches ~ 0.4 $\Delta$ at the van der Waals interfaces of NbSe$_2$ on graphene [36] or WTe$_2$ [7], and ~ 0.76 $\Delta$ in the case of epitaxial Al on InAs [6]. Here, $\Delta'$ induced by NbN is but a few percent of its parent gap. Fermi velocity mismatch is unlikely to be a major source since the Fermi velocity in NbN is of the same order as in graphene [51]. Interfacial disorder or tunnel barrier can also lower interface transparency [47]. However, our NbN/Pd/graphene interface is very smooth (RMS roughness of 1.8 Å) and has a low contact resistance of ~ 130 Ω · μm, comparable to other SC/graphene interfaces [12, 19, 22, 24, 36]. This seems to suggest that direct interface disorder and tunnel barrier are not main sources. What other mechanism then, can limit $\Delta'$ in our experiment? We speculate that disorder in the SC contact plays an important role. Our NbN films are strongly disordered given its relatively low $T_c \approx 7$ K, in comparison to a typical literature value higher than 10 K [2, 5, 19, 51]. The normal resistivity of our as-sputtered NbN is ~ 19 μΩ·m, which is consistent with disordered NbN films with a reduced $T_c$, but substantially larger than ~1 μΩ·m for a high-quality NbN film [51]. In addition, the magnetic interference pattern shown in Fig. 3c is rapidly suppressed by increasing $B$ and the supercurrent of the JJ is strongly dependent upon the history of the applied $B$-field or current bias, suggesting vortex trapping induced by disorders created in the sputtering process [51, 52] (See Supplementary Section 10). In 3D SN junctions, a strongly disordered SC may reduce the induced gap in the normal metal through $\Delta_N \approx \Delta_S \tau_S / (\tau_S + \tau_N)$, that is, if the quasiparticle relaxation time in the SC, $\tau_S$, is much smaller than its counterpart in the normal metal, $\tau_N$, then the induced gap $\Delta_N$ is much smaller than the parent gap $\Delta_S$ [10, 53]. We suspect that a large "quality mismatch" between the NbN and graphene films used in our experiment may have contributed to the smallness of $\Delta'$. The use of SC films with higher quality, such as Nb/Au, may substantially enhance the induced gap and the superconducting performance of the via contact.

In summary, we have realized tightly conforming, low-resistance, van der Waals-like interfaces between a 3D SC contact and a 2D material using a via contact approach and demonstrated Josephson current in NbN/Pd-graphene junctions. This planar contact geometry can be understood using an effective induced gap in the contact region. The induced gap is several tens of μeV in our experiment. We discuss potential factors impacting its magnitude and future directions for improvement. The via contact method can be generalized to many 3D SCs and is ideally suited to work with air-sensitive and damage-prone surfaces, e.g. topological insulators such as $Bi_2Se_3$ (See Supplementary Fig. S11). We hope that our work will stimulate more future effort in making novel superconducting heterostructures using the via method.

**Acknowledgements**

C.L., Y.L., K.D.H, J.L.T, D.R.H, M.K., and J.Z. are supported by the Penn State MRSEC for Nanoscale Science (DMR-2011839) and acknowledge the use of the center's low-temperature transport facilities (DOI: 10.60551/rxfx-9h58). The $Bi_2Se_3$ single crystal growth work was performed at the Pennsylvania State University Two-Dimensional Crystal Consortium–Materials Innovation Platform (2DCC-MIP), which is supported by NSF Cooperative Agreement No. DMR-2039351. K.D.H., J.L.T., and D.R.H. also acknowledge support through startup funds from the Penn State Eberly College of Science, Department of Chemistry, College of Earth and Mineral Sciences, Department of Materials Science and Engineering, and Materials Research Institute. The authors acknowledge the use of the PSU Materials Characterization Lab. The authors also acknowledge Arashdeep S. Thind and Robert F. Klie for access and support at the University of Illinois Chicago Electron Microscopy Core. K.W. and T.T. acknowledge support from the JSPS KAKENHI (Grant Numbers 21H05233 and 23H02052), the CREST (JPMJCR24A5), JST and World Premier International Research Center Initiative (WPI), MEXT, Japan. We thank Du Xu, Peng Wei, Javad Shabani, and Ivan V. Borzenets for helpful discussions and Ruoxi Zhang for the LabVIEW program used in measurements.

**Author contributions:**

C.L. and J.Z. designed the experiment. C.L. fabricated the devices. C.L. and Y.L. made the transport measurements under the supervision of M.K. and J.Z.; K.D.H and J.L.T. performed the FIB and STEM measurements under the supervision of D.R.H; Y.G. synthesized the $Bi_2Se_3$ crystals under the supervision of Z.M.; K.W. and T.T. synthesized the h-BN crystals; C.L. and J.Z. analyzed data; C.L. and J.Z. wrote the manuscript with input from all authors.

**Competing interests:**

The authors declare no competing interests.

**Methods**

**Device stacking and fabrication**
See Supplementary Information Section 1.

**Microscopy and elemental analysis**
Electron-transparent cross-sectional lamellae for ADF-STEM analysis were prepared using an FEI Helios dual-beam $Ga^+$ focused ion beam (FIB). Prior to FIB preparation, ~15 nm of amorphous carbon was deposited onto the samples in a Leica sputter coater. The lamellae were extracted from the NbN/Pd contact regions and were thinned using FIB accelerating voltages of 30 kV and 5 kV, followed by a final cleaning step at 2 kV. ADF-STEM imaging was performed on a dual spherical aberration-corrected FEI Titan[3] G2 60-300 S/TEM at an accelerating voltage of 300 kV with a convergence angle of 25.2 mrad, with ADF detector collection angles of 42–244 mrad. High-resolution TEM and EDX elemental analysis were performed on an FEI Talos F200X at an accelerating voltage of 200 kV. EELS mapping and additional ADF-STEM imaging were collected on a probe-corrected JEOL JEM-ARM200CF S/TEM at an accelerating voltage of 200 kV with a convergence angle of 30 mrad and STEM detector collection angles of 40–160 mrad. EELS was performed using a Continuum Gatan Imaging Filter (GIF) spectrometer with an energy dispersion of 0.30 eV/ch to observe carbon, nitrogen, and oxygen edge signals.

**Transport measurements**
Measurements were performed in a cryogen-free dilution refrigerator with a base temperature of ~ 10 mK and in a He-3 cryostat with a base temperature of 320 mK. To perform differential resistance measurements, we applied a dc current bias with a small ac modulation of 1 nA and measured the ac and dc voltage using a lock-in amplifier (Stanford Research SR860) and a dc voltmeter (Keysight 34401A Digital Multimeter), respectively. NbN/Pd films were characterized in a Quantum Design Dynacool system with a magnet of 14 T and a base temperature of ~ 2 K. Resistances were measured with an excitation current of 1 µA.

**Data availability:** The data needed to reproduce the main text and supplementary information figures are available from the corresponding author upon reasonable request.

**Code availability:** The codes used in theoretical simulations and calculations are available from the corresponding author upon reasonable request.

**Main text figure 1 (two column figure)**

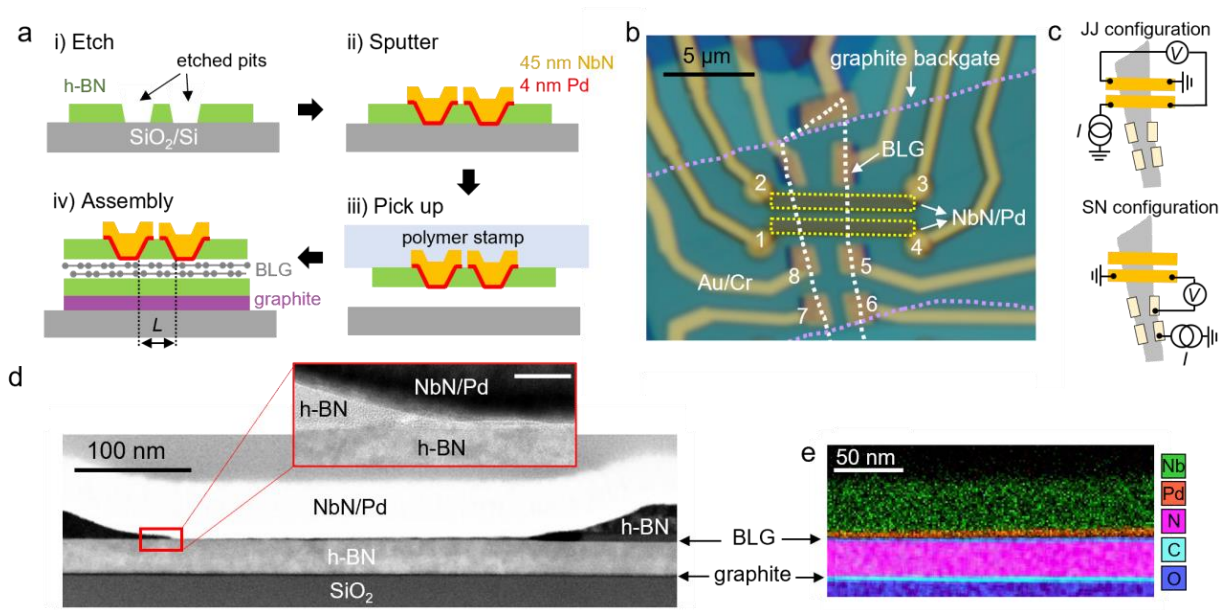

**Fig. 1 | Fabrication of NbN/Pd via contacts and device structure. a**, Schematics of the via contact fabrication and assembly process. **b**, Optical image of device J008. The white and purple dashed lines outline the BLG flake and the graphite backgate respectively. The BLG channel between the two NbN/Pd strips measures 2.9 μm × 0.69 μm . Au/Cr electrodes are made to the NbN/Pd strips (1–4) and BLG (5–8). **c**, Measurement configurations for a JJ and an SN junction. **d,** Cross-sectional ADF-STEM image of a junction cut along the length of the BLG flake. The inset is a zoomed-in TEM image corresponding to the region marked in the red box, and the scale bar is 20 nm. **e,** Overlaid EDX/EELS maps taken in the center of one NbN/Pd/BLG contact (Nb and Pd are EDX signals, and N, C, and O are EELS signals).

**Main text figure 2 (one column figure)**

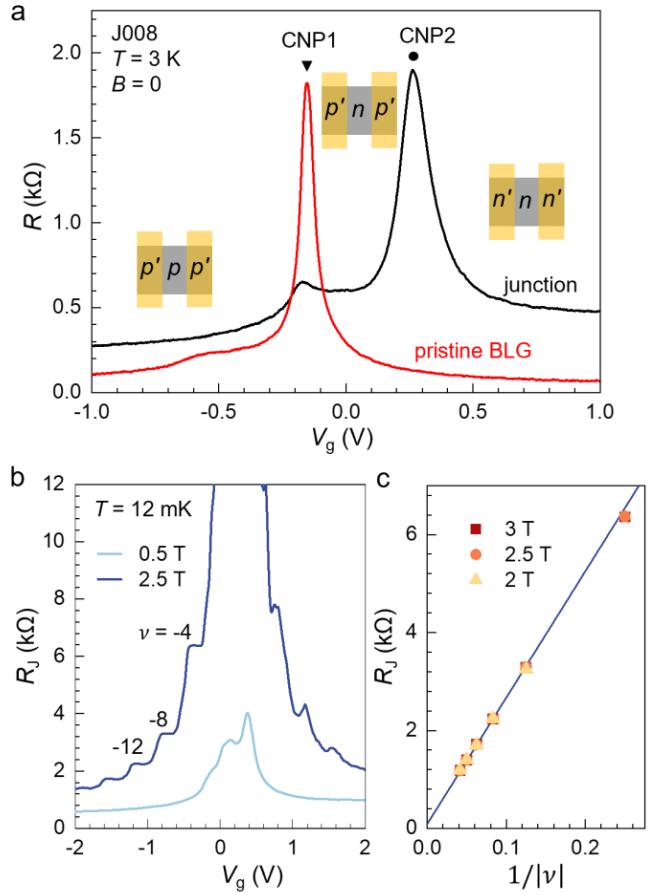

**Fig. 2 | Normal transport of the NbN/Pd-BLG Josephson junction. a,** $V_g$-dependent junction resistance $R_J = R_{1,3;4,2}$ (black trace) obtained using the JJ configuration in Fig. 1c, and $V_g$-dependent four-terminal resistance of the pristine BLG $R_{7,8;6,5}$ (red trace). $T = 3$ K. The insets illustrate the carrier density profile in the BLG for the black trace. **b,** Junction resistance $R_J$ versus $V_g$ at selected magnetic fields as labeled. At $B > 2$ T, $R_J(V_g)$ develops quantum Hall plateaus on the hole side, a few of which are labeled. The quantum Hall effect allows us to determine the gating efficiency of the backgate to be $6.25 \times 10^{11}$ V$^{-1}$cm$^{-2}$. Measurements at more fields are given in Fig. S4 of the Supplementary Information. **c,** $R_J$ at the quantum Hall resistance plateaus versus the inverse filling factor $1/\nu$ for the hole side. The blue solid line is a linear fit to Eq. 1, from which we extract $R_C = 45 \pm 4$ Ω for one NbN/Pd contact. $R_C$ offers a good estimate of the $B = 0$ contact resistance when it is small and $B$-independent. Data are from device J008.

**Main text figure 3 (two column figure)**

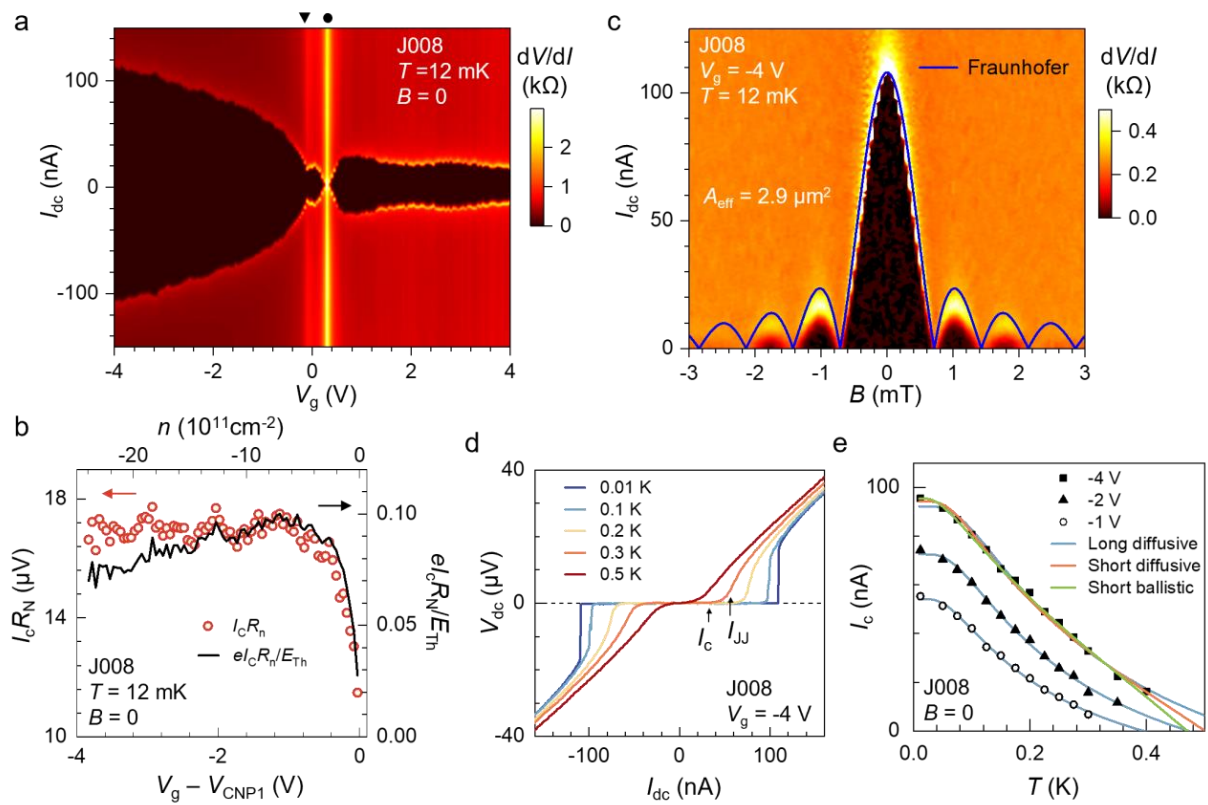

**Fig. 3 | Josephson effect in NbN/Pd-BLG junctions. a,** False-colored map of differential resistance $dV/dI$ versus the dc current bias $I_{dc}$ and backgate voltage $V_g$. **b,** Extracted $I_c R_N$ (red circles) and $I_c R_N/E_{Th}$ (black solid line) as a function of $V_g$. **c,** Map of $dV/dI$ versus $I_{dc}$ and magnetic field $B$ obtained on J008. **d,** Temperature-dependent $V$-$I$ characteristics taken at $V_g = -4$ V and selected temperatures as labeled. The critical current $I_c$ is defined as the current corresponding to a small threshold voltage $V = 0.5$ μV. $I_{JJ}$ is defined as the current corresponding to the steepest slope of the $V$-$I$ curve, i.e., the peak of $dV/dI$. The arrows point to the $I_c$ and $I_{JJ}$ values for the 0.3 K trace. **e,** Extracted $I_c$ versus $T$ at different $V_g$'s. The blue solid lines are fits to the long diffusive model. Fitting results of $\alpha_{1,2}$ and $b$ are given in Supplementary Fig. S8. The orange and green solid lines represent the short diffusive and short ballistic fits to the $V_g = -4$ V trace, respectively. Data are from device J008.

**Main text figure 4 (one column figure)**

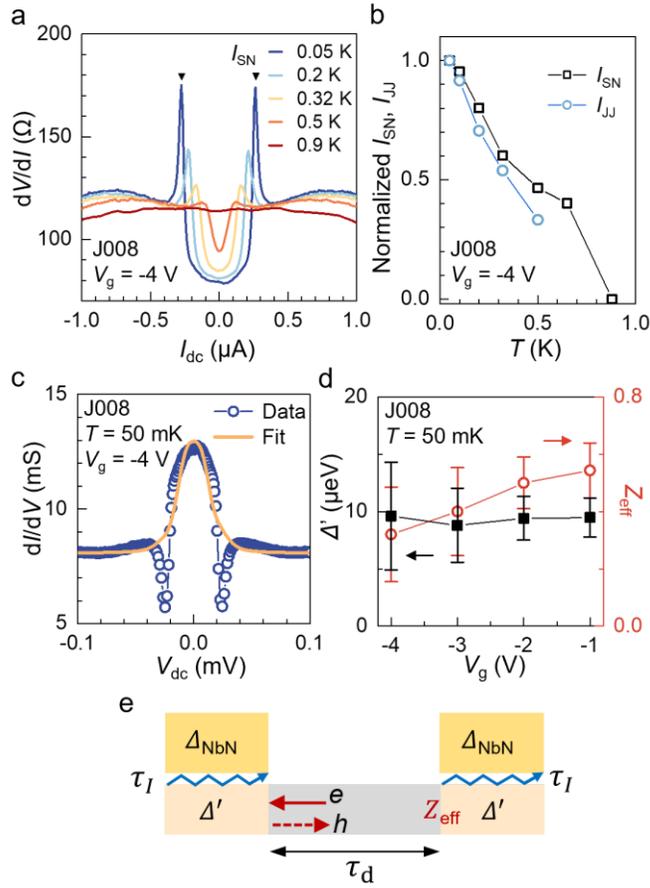

**Fig. 4 | Andreev reflections in a single NbN/Pd-BLG SN junction. a,** Differential resistance $dV/dI$ versus $I_{dc}$ taken at $V_g = -4$ V and selected temperatures as labeled on an SN junction in J008. The zero-bias resistance reduction is approximately 1/3 at $T = 0.05$ K. **b,** Temperature dependence of the normalized $I_{SN}$ and $I_{JJ}$ using data in **a** and Fig. 3d. $I_{SN}$ and $I_{JJ}$ are normalized to their respective 50 mK values. **c,** Differential conductance $dI/dV$ versus $V_{dc}$, obtained by integrating the corresponding raw data in **a** and then taking $dI/dV$. The orange solid line is the BTK fit to the data. **d,** Extracted gap $\Delta'$ and the effective barrier strength $Z_{eff}$ from the fits at selected $V_g$'s. **e,** Schematic of an effective gap model of our JJ. The proximity effect induces an effective gap of $\Delta'$ in the BLG region underneath the contact, which leads to an interface dwell time $\tau_I = \hbar/\Delta'$. $\tau_d = \hbar/E_{Th}$ represents the dwell time in the normal BLG channel. In our devices, $\tau_I \gg \tau_d$, supporting the application of the effective gap model.

# Supplementary Information:

# Building 3D superconductor-based Josephson junctions using a via transfer approach

Cequn Li[1#], Le Yi[1,2], Kalana D. Halanayake[3], Jessica L. Thompson[3], Yingdong Guan[1,4], Kenji Watanabe[5], Takashi Taniguchi[6], Zhiqiang Mao[1,4], Danielle Reifsnyder Hickey[2,3,7], Morteza Kayyalha[2,8], Jun Zhu[1,2*]

1. Department of Physics, The Pennsylvania State University, University Park, PA, USA
2. Materials Research Institute, The Pennsylvania State University, University Park, PA 16802, USA
3. Department of Chemistry, The Pennsylvania State University, University Park, PA 16802, USA
4. 2-Dimensional Crystal Consortium, The Pennsylvania State University, University Park, PA 16802, USA
5. Research Center for Electronic and Optical Materials, National Institute for Materials Science, 1-1 Namiki, Tsukuba 305-0044, Japan
6. Research Center for Materials Nanoarchitectonics, National Institute for Materials Science, 1-1 Namiki, Tsukuba 305-0044, Japan
7. Department of Materials Science and Engineering, The Pennsylvania State University, University Park, PA 16802, USA
8. Department of Electrical Engineering, The Pennsylvania State University, University Park, Pennsylvania 16802, USA

# Present address: Laboratory of Atomic and Solid State Physics, Cornell University, Ithaca, NY 14853, USA

* Correspondence to: jxz26@psu.edu (J. Zhu)

**Contents**

1. Fabrication of NbN/Pd via contacts and graphene Josephson junctions

We describe the steps to fabricate NbN/Pd via contacts and van der Waals Josephson junctions (JJ). First, we exfoliated h-BN flakes on the SiO$_2$/Si substrates using Scotch tape. To remove the polymer residue on h-BN, we performed O$_2$/Ar annealing (75/750 sccm, 450 ℃, 3 hours) in a tube furnace. We selected ~ 20–30 nm thick h-BN flakes under an optical microscope and examined their thickness and surface morphology using an atomic force microscope (AFM). The well-controlled thickness and uniformity of h-BN ensured a uniform etching of the underneath SiO$_2$/Si substrate for low surface roughness. We used standard electron beam lithography (EBL) and reactive ion etching (RIE) (CHF$_3$ 40 sccm/O$_2$ 4 sccm, power 40 W) to open rectangular windows about 500 nm wide on the target h-BN flake. An example of etched pits on h-BN is shown in Fig. S1a. Next, we spin-coated a new layer of resist (PMMA 950A2/MMA EL6) and patterned NbN/Pd (45/4 nm) into the etched pits. We sputtered the Pd layer using RF magnetron sputtering (Ar, pressure 5 mTorr, power 200 W) and NbN using DC magnetron sputtering (Nb target, Ar/N$_2$ 87/13%, pressure 7 mTorr, power 250 W) without breaking vacuum. During the sputtering, we kept the substrate right above the target and avoided substrate rotation to ensure deposition at a straight angle. To avoid the tall build-ups on the side of NbN, we sonicated the chip in an isopropyl alcohol (IPA) bath for only ~7s and repeated this step 5 times after lift-off. The optical image of a NbN/Pd-embedded h-BN is shown in Fig. S1b.

Next, we assembled the heterostructure by transferring NbN/Pd via contacts, monolayer graphene (MLG) or bilayer graphene (BLG), h-BN, and graphite backgate sequentially using a Poly(Bisphenol A carbonate) (PC) stamp. The transfer process was performed inside an Ar-filled glove box to prevent oxidation. A complete stack is shown in Fig. S1c. Finally, we used EBL and electron-beam evaporation to pattern electrodes, in which Au/Ti (60/5 nm) was made to contact with NbN, followed by exposing the edge of the graphene using RIE and patterning Au/Cr (60/5 nm) on both NbN and the exposed graphene edge (see Fig. S1d).

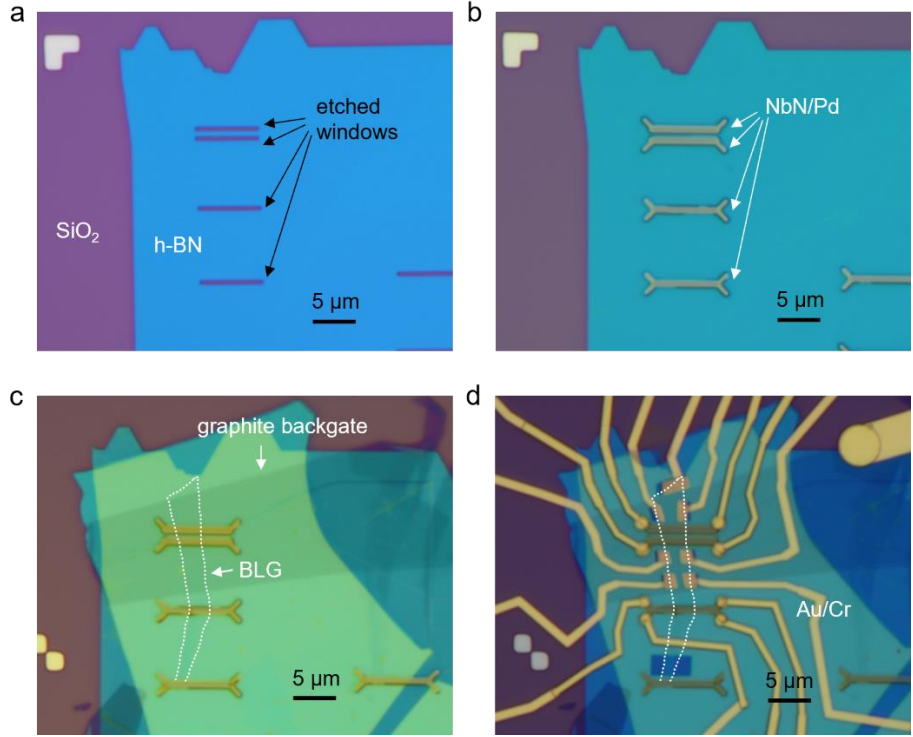

**Fig. S1 | Fabrication of NbN/Pd via contacts and graphene Josephson junctions. a,** Optical image of an h-BN flake with etched windows. **b,** NbN/Pd (45/4 nm) sputtered into the etched windows. **c,** A complete NbN/Pd/BLG/h-BN/graphite stack on SiO$_2$/Si substrate. **d,** Optical image of a complete device J008.

## 2. Characterization of NbN/Pd thin films

We perform four-terminal transport measurements on the as-sputtered NbN/Pd films. Figure S2a shows the temperature dependence of the resistance in films from different batches. Overall, we find the superconducting transition temperature $T_c$ varies from 6 to 11 K. We note that the quality of the films is very sensitive to the sputtering chamber conditions such as the base pressure, water and oxygen concentrations, and the N$_2$/Ar ratio. Figure S2b shows the magnetic field dependence of the resistance in film E, where the upper critical field $H_{c2}$ is much larger than 14 T at $T < 8$ K. Most of our films remain zero resistance up to 14 T at 2 K. We also characterize the longitudinal resistance $R_{xx}$ as a function of the current bias measured on a 5 μm-wide NbN/Pd Hall-bar device (Fig. S2c), where we obtain the critical current density $J_c$ of ~ 400 μA/μm across the width of the device.

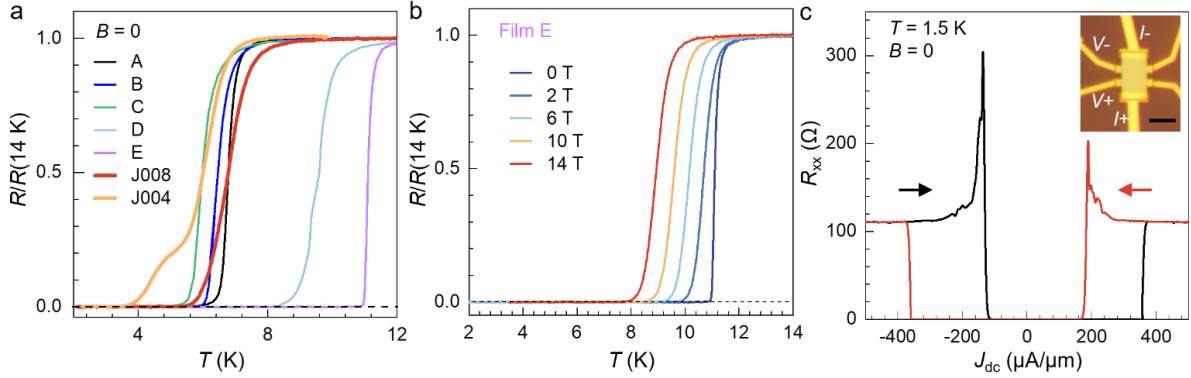

**Fig. S2 | Superconductivity of NbN/Pd thin films. a**, Normalized sample resistance versus temperature. Sample A–E are films sputtered from different runs. In comparison, red and yellow curves are measured on the ~500 nm-wide NbN/Pd control stripe on device J008 and J004, respectively. **b**, Temperature-dependence of sample resistance at different perpendicular magnetic fields obtained on the film E. **c**, Longitudinal resistance $R_{xx}$ as a function of current density $J_c$ measured on a 5 µm-wide NbN/Pd Hall-bar device. The inset shows the measurement configuration. Current source ($I+$), drain ($I-$), and high/low voltage ($V+/-$) probes are labeled. Scale bar, 5 µm.

## 3. Characterization of the structure of NbN/Pd via contacts

Figure S3a shows a false-colored large-scale cross-sectional annular dark-field (ADF-) scanning transmission electron microscope (STEM) image of a representative NbN/Pd-BLG JJ device J009 whose structure is similar to J008 studied in the main text. The build-ups on the two sides of NbN are present but too small to short the adjacent contacts. A zoomed-in ADF-STEM image of a single junction is presented in Fig. S3b, adopted from Fig. 1d of the main text. The energy-dispersive X-ray spectroscopy (EDX) and electron energy-loss spectroscopy (EELS) maps verify that NbN/Pd is flush with the h-BN, and no voids are observed at the bottom of the interface, as shown in Fig. S3c. No deterioration or contamination is observed on the BLG or at the interface. We also study the surface morphology of the NbN/Pd via contacts using AFM in Fig. S3e–f. To expose the Pd side of a NbN/Pd contact to AFM, we picked up one NbN/Pd contact using a PC stamp and then peeled off the PC film. The PC film was flipped and deposited on a clean Si substrate for AFM imaging, as illustrated in Fig. S3d. The bottom surface of the Pd layer is quite flat as it is templated by the $SiO_2$ surface during the sputtering, with a root-mean-square roughness of 1.8 Å (Figure S3f), consistent with the previous study [1]. In comparison, the top surface of NbN is much rougher. The flat bottom surface of the vias is key to the planar and smooth NbN/Pd-graphene interface.

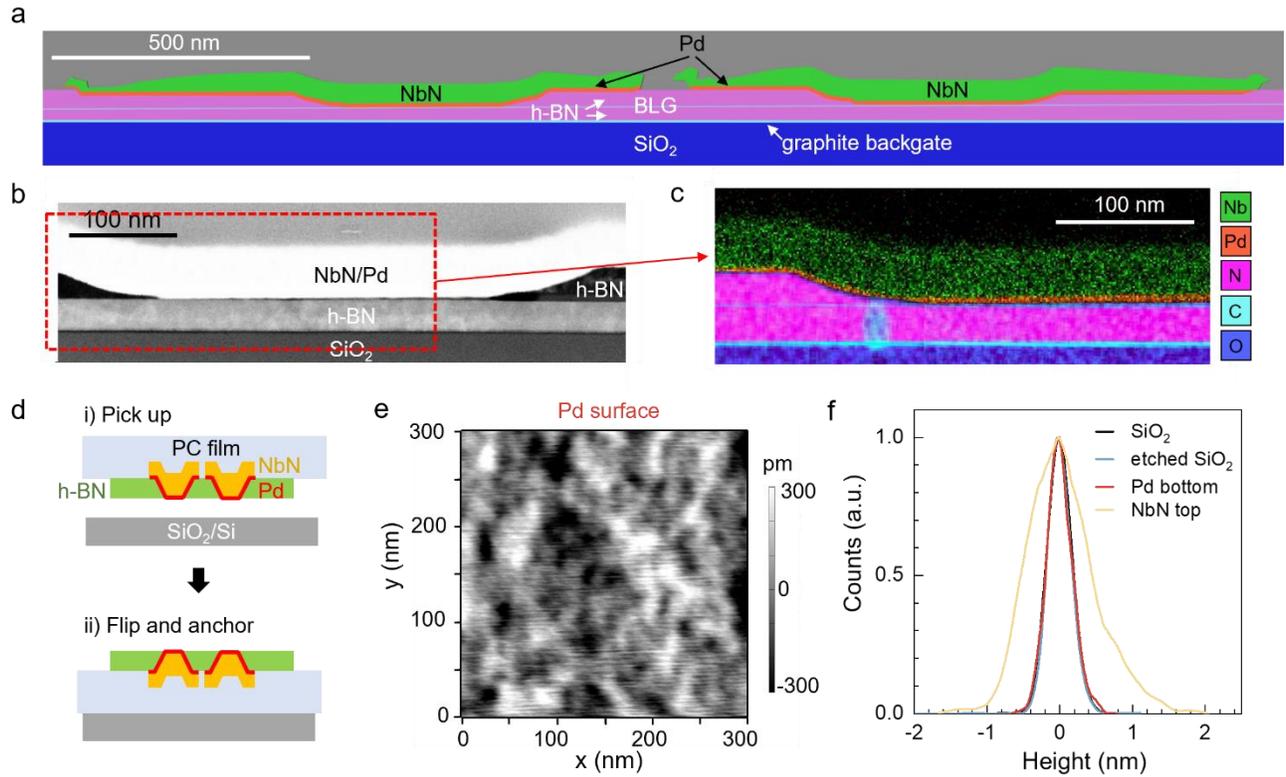

**Fig. S3 | Structure and surface morphology of NbN/Pd via contacts. a,** False-colored cross-sectional ADF-STEM image of a NbN/Pd/BLG JJ. **b,** Cross-sectional STEM image of a single NbN/Pd/BLG junction. **c,** Overlaid EDX/EELS maps corresponding to the region marked in the red box in **b** (Nb and Pd are EDX signals, and N, C, and O are EELS signals). The cyan patch is due to electron beam carbon contamination. **d,** Schematics of picking up and inverting a NbN/Pd contact for imaging the Pd side. **e,** AFM scan of the bottom side of a NbN/Pd via contact. **f,** Height histograms for bare and etched SiO$_2$, Pd bottom, and NbN top surface roughness. The surface roughness is the standard deviation of the height histogram generated from ~300 nm × 300 nm square AFM area.

## 4. Additional data on the normal transport measurements of J008

Figure S4a shows the Landau fan diagram of device J008 made on BLG. We observe resistance plateaus with filling factor $\nu = \pm 4, \pm 8, \pm 12$ ..., which are expected for BLG. These filling factors follow the Streda formula $\nu = nh/eB$ (where $n$ is the carrier density, h is the Planck constant, $e$ is the electron charge, and $B$ is the magnetic field) (Refs. [2-4]), as indicated by the black dashed lines. The two sets of Landau fans emanate from CNP 1 and CNP 2, further supporting the carrier density profile illustrated in Fig. 2a of the main text. Figure S4b plots the quantum Hall states in J008 at selected magnetic fields for the extraction of contact resistance as described in the main text [5].

Figure S4c plots the mean free path $L_{\text{mfp}}$ of the graphene channel estimated using $L_{\text{mfp}} = h/(2e^2 k_F R_{\text{sheet}})$, where $k_F = \sqrt{\pi n}$ is the Fermi wave vector, $n$ is the carrier density, and $R_{\text{sheet}}$ is the graphene sheet resistance given by $R_{\text{sheet}} = (R_J - 2R_C) \times W/L$ [6]. $2R_C \approx 100\ \Omega$ and $660\ \Omega$ respectively in J008 and J004. The mean free path on the hole side is generally 100–200 nm, smaller than $L$, which places the junction in the intermediate to diffusive regime. Because the determination of $L_{\text{mfp}}$ sensitively depends on the accurate knowledge of $R_C$, we have also considered the scenario where the graphene channel approaches the ballistic limit (See the short ballistic fit in Fig. 3e of the main text). The electron side is more complicated due to the doping inhomogeneity in the NbN/Pd/graphene contact area which dominates the resistance. Therefore, we do not discuss the mean free path on the electron side.

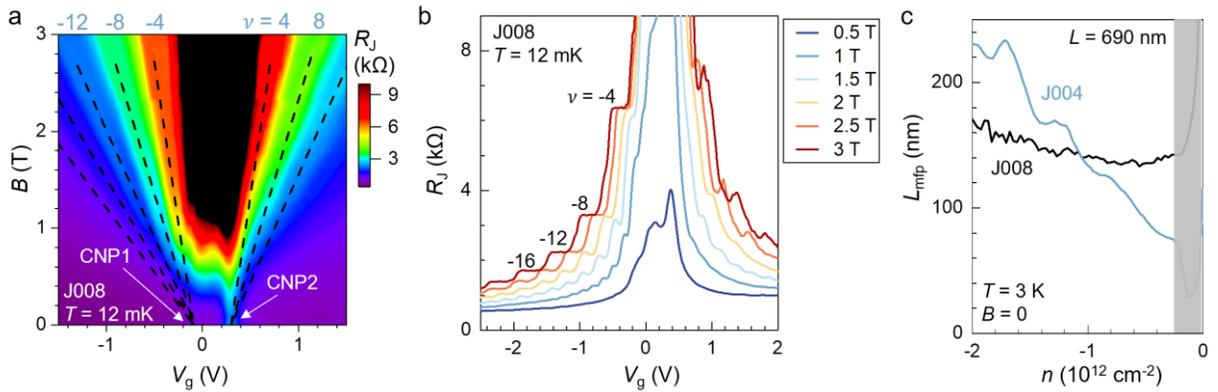

**Fig. S4 | Quantum Hall states and the estimation of the graphene mean free path. a,** A color map of the junction resistance $R_J$ versus the backgate voltage $V_g$ and magnetic field $B$ showing Landau levels emanating from the CNPs of the channel and the contact area respectively. From device J008. **b,** $R_J(V_g)$ traces at selected magnetic fields. **c,** Estimated mean free path $L_{\text{mfp}}$ of the graphene channel in the junction as a function of hole density in device J008 and J004. Here we used $R_J(V_g)$ at $T = 3$ K instead of $R_N(V_g)$ to calculate $L_{\text{mfp}}$, which leads to an underestimation of $L_{\text{mfp}}$ by about 30%.

## 5. Extraction of the gap of NbN using the differential conductance

Figure S5 shows the differential conductance $dI/dV$ as a function of dc voltage bias $V_{dc}$ obtained on device J008 at selected backgate voltages $V_g$. The conductance peak at $eV_{dc} = 0.9$ meV is present for all $V_g$ and we attribute it to the superconducting gap of NbN $\Delta_{NbN}$, which agrees with the Bardeen–Cooper–Schrieffer (BCS) gap $1.76 k_B T_c$ where $T_c = 6.7$ K is obtained on a NbN/Pd control stripe next to the JJ. The conductance suppression at $eV_{dc} < \Delta_{NbN}$ indicates a low interface transparency between NbN/Pd and graphene.

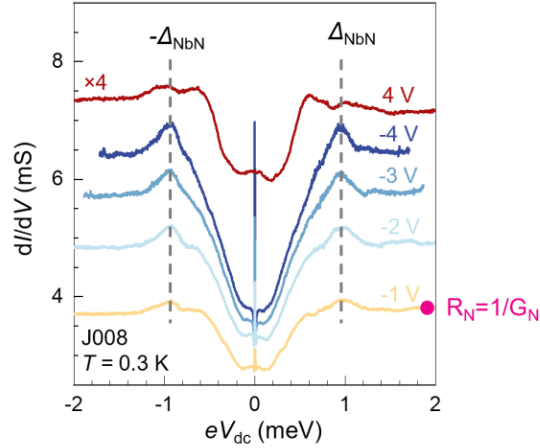

**Fig. S5 | Extraction of the gap of NbN.** Differential conductance $dI/dV$ versus $V_{dc}$ measured at $T = 0.3$ K at selected gate voltages $V_g$ in J008. The $dI/dV$ for $V_g = 4$ V is multiplied by 4 for clarity. The zero-bias conductance remains finite because of the suppression of the supercurrent by the large bias applied in these measurements (see further discussions in Section 10). $\Delta_{NbN} = 0.9$ meV is determined by the location of the conductance peak. The normal state resistance $R_N$ is the inverse of the saturated differential conductance at large bias, e.g., the magenta dot on the yellow curve.

## 6. Fit to the effective short diffusive and short ballistic model

For a small weak link in the diffusive limit ($L_{mfp} \ll L$), the temperature dependence of the supercurrent can be well described by the Kulik-Omelyanchuk (KO-1) theory [7], expressed as

$$I(T, \varphi) = \frac{2\pi k_B T}{e R_N} \times \sum_{\omega > 0} \frac{2\gamma \cos\left(\frac{\varphi}{2}\right)}{\delta} \arctan \frac{\gamma \sin\left(\frac{\varphi}{2}\right)}{\delta}, \quad \text{(Eq. S1)}$$

where $\gamma$ is the $T$-dependent superconducting gap, $\omega = \pi k_B T (2m + 1)/\hbar$ is the Matsubara frequency with $m = 0, 1, 2 \ldots$, and $\delta = \sqrt{\gamma^2 \cos^2(\varphi/2) + (\hbar \omega)^2}$. The phase difference between the two superconducting electrodes $\varphi$ is treated as a variable to maximize Eq. S1 to determine the critical current $I_c$ at a given $T$. In the $T = 0$ limit, Eq. S1 converges and $eI_cR_N$ saturates at approximately $2.07\gamma_0$, where $\gamma_0 = \gamma(T = 0)$.

We fit $I_c(T)$ in Fig. 3e of the main text using Eq. S1, following the prior study in the short diffusive graphene JJs [8]. To avoid the potential pre-factor $\alpha < 1$ on the right-hand side of Eq. S1 [8], the normalized $I_c$ is plotted and fitted as shown in Fig. S6. We first approximate the temperature dependence of $\gamma$ using $\gamma(T) = \gamma_0 \tanh(1.74\sqrt{T_c/T - 1})$, where $T_c = 6.7$ K is the critical temperature of NbN. We fit the induced gap $\gamma_0$ which is a fraction of the parent gap of NbN, following a similar strategy for fitting the induced gap in Nb/HgTe [9]. A representative fit to the data obtained at $V_g = -4$ V in J008 is given by the blue solid line in Fig. S6. The obtained $\gamma_0$ at different $V_g$'s is in the range of 20–30 μeV (see Table S1), consistent with the energy scales extracted from the $dV/dI$ measurements of a single SN junction. However, the fit deviates from the data at approximately $T > \gamma_0/k_B \approx 0.3$ K, possibly due to the suppression of $\gamma$ at high temperatures. Indeed, Fig. 4 of the main text shows that the Andreev reflections already vanish in the SN junction at approximately $T = 1$ K. These findings suggest that $\gamma$ may be subject to a much smaller $T_c$ compared to $\Delta_{NbN}$.

Therefore, we adopt an effective gap function $\gamma(T) = \gamma_0 \tanh(1.74\sqrt{T^*/T - 1})$ and fit both $\gamma_0$ and the effective critical temperature $T^*$ using Eq. S1. This approach results in a better fit to the data given by the orange curve in Fig. S6, also plotted in Fig. 3e of the main text. We obtain $\gamma_0$ in the range of 23–33 μeV and $T^*$ in the range of 0.35–0.5 K. Although $T^*$ is an order of magnitude smaller than the $T_c$ of NbN, the fitting of $\gamma_0$ remains robust since it is primarily governed by the $I_c(T)$ behavior near zero temperature. The ratio $eI_cR_N/\gamma_0$ is about 0.5–0.7, smaller than $2.07$ predicted by the KO-1 theory. This could indicate that our JJ departs from the diffusive junction limit ($L_{mfp} \ll L$). Table S1 summarizes the fitting results obtained in J008.

We have also considered the scenario where the graphene channel may be approaching the ballistic limit in the heavy hole regime and modeled the JJ using the short ballistic theory [10] given by

$$I_c(T,\varphi) = \frac{\pi\gamma}{2eR_N} \frac{\sin\varphi}{\sqrt{1 - \mathcal{T}\sin^2\left(\frac{\varphi}{2}\right)}} \tanh\left(\frac{\gamma}{2k_BT}\sqrt{1 - \mathcal{T}\sin^2\left(\frac{\varphi}{2}\right)}\right) \quad \text{(Eq. S2)}$$

where $\mathcal{T} \in [0,1]$ is the transmission coefficient of the S-N interface. Here, $\gamma(T)$ is an effective gap function similar to what's described in the short diffusive mode (Eq. S1). The green solid line in Fig. S6 represents a fit to Eq. S2 for the $V_g = -4$ V data. This fit is also shown in Fig. 3e of the main text. Equation S2 fits all our data in the hole regime well. The fitting results of $\gamma$ and $\mathcal{T}$ at different $V_g$'s are summarized in Table S1.

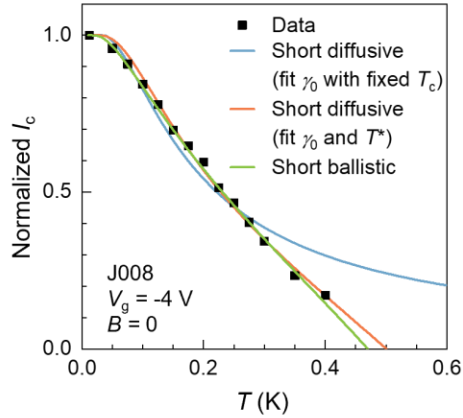

**Fig. S6 | Short diffusive and short ballistic fits to the data.** Normalized $I_c$ versus $T$ at different $V_g$'s in J008. The blue solid line is the short diffusive fit to Eq. S1 using $\gamma(T) = \gamma_0 \tanh\left(1.74\sqrt{T_c/T - 1}\right)$ with $T_c = 6.7$ K. The orange solid line plots the short diffusive fit using $\gamma(T) = \gamma_0 \tanh\left(1.74\sqrt{T^*/T - 1}\right)$ to fit both $\gamma_0$ and $T^*$. The green solid line plots the short ballistic fit to fit $\gamma_0$, $\mathcal{T}$, and $T^*$. $T^* \ll T_c$.

|  |  |  | $V_g = -4$ V | $V_g = -3$ V | $V_g = -2$ V | $V_g = -1$ V |
|---|---|---|---|---|---|---|
| Short diffusive fit | Fit $\gamma_0$ with fixed $T_c = 6.7$ K | $\gamma_0$ (μeV) | $29 \pm 2$ | $27 \pm 2$ | $24 \pm 2$ | $20 \pm 2$ |
|  | Fit $\gamma_0$ and $T^*$ | $\gamma_0$ (μeV) | $33 \pm 3$ | $30 \pm 3$ | $27 \pm 3$ | $23 \pm 2$ |
|  |  | $T^*$ (K) | $0.49 \pm 0.03$ | $0.48 \pm 0.02$ | $0.42 \pm 0.02$ | $0.35 \pm 0.03$ |
| Short ballistic fit | Fit $\gamma_0$, $\mathcal{T}$, and $T^*$ | $\gamma_0$ (μeV) | $49 \pm 5$ | $40 \pm 10$ | $34 \pm 10$ | $28 \pm 6$ |
|  |  | $\mathcal{T}$ | $0.99 \pm 0.01$ | $0.97 \pm 0.03$ | $0.93 \pm 0.07$ | $0.92 \pm 0.08$ |
|  |  | $T^*$ (K) | $0.47 \pm 0.02$ | $0.47 \pm 0.02$ | $0.41 \pm 0.02$ | $0.35 \pm 0.02$ |

**Table S1 | The zero-temperature induced gap $\gamma_0$, interface transmission $\mathcal{T}$, and effective critical temperature $T^*$ at selected $V_g$'s in J008 obtained from the various fits to the curves in Fig. 3e of the main text.**

## 7. Josephson effect in device J004 made on monolayer graphene/h-BN superlattice

Figure S7a shows an optical image of device J004 with the width $W = 1.4$ µm and length $L = 0.69$ µm, where the MLG is aligned with the top h-BN at a small misalignment angle. Indeed, we observe two charge neutrality points (CNP) in the normal resistance (Fig. S7b), the main CNP of the MLG at $V_g = 0.14$ V, and the satellite CNP due to the miniband structure of the MLG/h-BN superlattice at $V_g = -5.8$ V. The carrier density corresponding to the satellite CNP is $n_{sp} = -3.8 \times 10^{12}$ cm$^{-2}$. The superlattice wavelength $\lambda$ is calculated using $\lambda = \sqrt{\frac{4\pi}{3n_{sp}}} = 10.5$ nm. This misalignment angle $\theta$ can be estimated using the following formula [11]

$$\lambda = \frac{a}{\sqrt{\delta^2 + \theta^2}}, \qquad (\text{Eq. S2})$$

where the lattice constant $a = 2.46$ Å and lattice mismatch $\delta = 1.8\%$. In our sample, we estimate $\theta \approx 0.86°$.

Figure S7c plots the differential resistance $dV/dI$ as a function of the dc current bias $I_{dc}$ and gate voltage $V_g$ at the base temperature. $I_c$ vanishes at the main and satellite CNPs, with a maximum of ~30 nA. Figure S7d shows the Fraunhofer pattern obtained at $V_g = -2.5$ V, which is well described by a uniform current distribution. We extract an effective area $A_{eff} = 1.3$ µm² from the oscillation period. Using the same empirical formula $A_{eff} = W \times (L + 2L_{S,half})$ [2], we estimate $2L_{S,half} = 0.24$ µm, in good agreement with the measured length of the NbN/Pd electrode ~0.3 µm.

Figure S7e shows the temperature-dependent current-voltage ($I$-$V$) curves in J004. The retrapping current $I_r$ is slightly smaller than the switching current $I_c$, indicating an underdamped JJ. The $I_c R_N$ product for $-2.5$ V $< V_g < -0.1$ V is 10–25 µV, consistent with the findings in J008. We fit $I_c(T)$ to the long diffusive JJ model (Eq. 3 in the main text, reproduced as Eq. S3 here for completeness) [12-15],

$$eI_c R_N = \alpha_1 E_{Th} \left[1 - b \exp\left(\frac{-\alpha_2 E_{Th}}{3.2 k_B T}\right)\right], \qquad (\text{Eq. S3})$$

where $E_{Th} = \hbar D/L^2$ is the Thouless energy and $\alpha_1$, $\alpha_2$, and $b$ are fitting parameters, as shown in Fig. S7f. Our best fits yield $\alpha_1 = 0.05 \pm 0.01$, $\alpha_2 = 0.16 \pm 0.11$, and $b = 2.07 \pm 1.7$, consistent with the findings in J008.

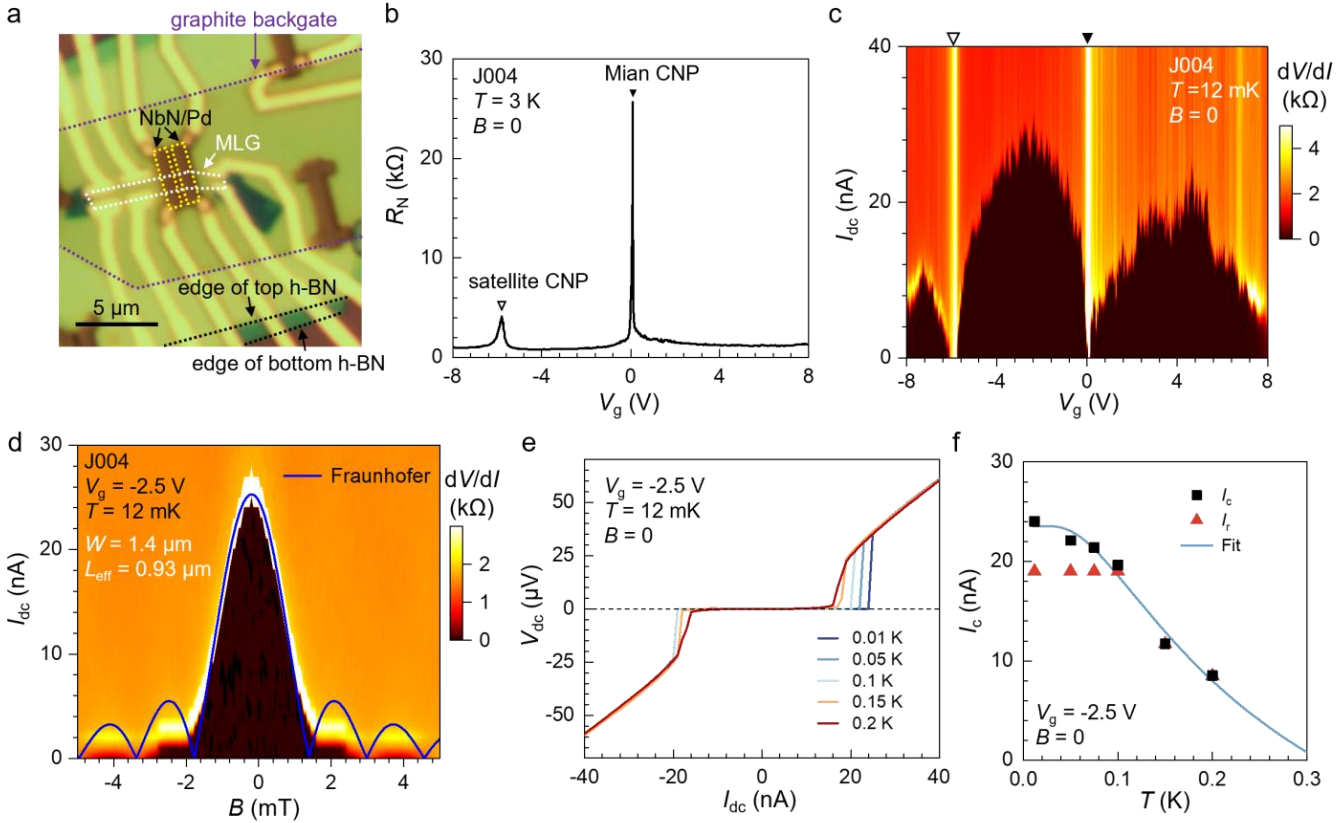

**Fig. S7 | Additional data on device J004. a**, Optical image of device J004. **b,** Junction resistance $R_J$ as a function of $V_g$ obtained at $T = 3$ K. **c,** Map of differential resistance $dV/dI$ vs $I_{dc}$ and $V_g$. **d,** Map of differential resistance $dV/dI$ vs current $I_{dc}$ and magnetic field $B$. **e**, $I$-$V$ curve obtained at $T = 12$ mK and $V_g = -2.5$ V. **f,** Extracted critical current $I_c$ and retrapping current $I_r$ versus $T$. $I_c$ is defined using a small voltage threshold $V_c = 0.1$ μV. The bule solid line is the long diffusive fit.

## 8. Comparison of $eI_cR_N/E_{Th}$ in long diffusive graphene JJs

Figure S8a plots $eI_cR_N/E_{Th}$ obtained in J008 and J004 at $T = 12$ mK together with the results reported for other long diffusive graphene JJs made using different SCs (Pb/Pd [16], Al/Ti [17],

ReW/Pd [17], Nb/Pd [17], MoRe [11], NbTiN [18], and NbSe$_2$ [15]). Our $eI_cR_N/E_{Th}$ is in the range of 0.03–0.25, on par with the findings in previous studies [13-18]. It also exhibits a clear trend of increasing $eI_cR_N/E_{Th}$ with increasing $L/\xi$, as the junction approaches the long junction limit. Figure S8b summarizes the fitting parameters $\alpha_1$, $\alpha_2$, and $b$ obtained using the long diffusive fits to the data shown in Fig. 3e of the main text.

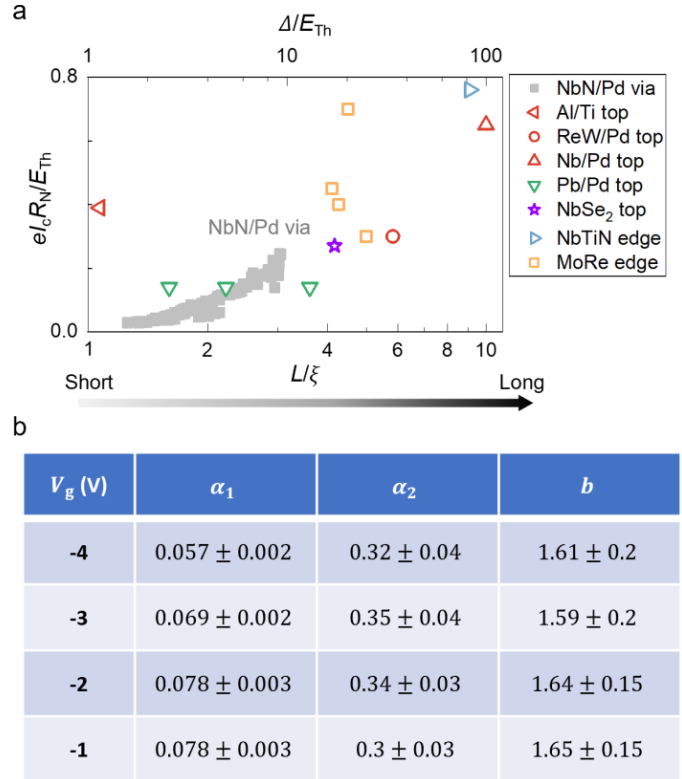

| $V_g$ (V) | $\alpha_1$ | $\alpha_2$ | $b$ |
|---|---|---|---|
| -4 | 0.057 ± 0.002 | 0.32 ± 0.04 | 1.61 ± 0.2 |
| -3 | 0.069 ± 0.002 | 0.35 ± 0.04 | 1.59 ± 0.2 |
| -2 | 0.078 ± 0.003 | 0.34 ± 0.03 | 1.64 ± 0.15 |
| -1 | 0.078 ± 0.003 | 0.3 ± 0.03 | 1.65 ± 0.15 |

**Fig. S8 | $eI_cR_N/E_{Th}$ in long diffusive graphene JJs. a,** Comparison of $eI_cR_N/E_{Th}$ obtained from J008 and J004 at different backgate voltages to other superconducting contacts made with different materials and styles as labeled. **b,** Table of $\alpha_1$, $\alpha_2$, and $b$ obtained from the long diffusive fits to the data at selected $V_g$'s in J008. Data and fits are given in Fig. 3e of the main text.

## 9. Andreev reflections in a single SN junction

Figure S9a shows a large-range $I_{dc}$ sweep of $dV/dI$ in a single SN junction in J008. We attribute the increase of $dV/dI$ in the small current bias range of $I_{dc} < 5$ μA to the single particle tunneling at the SN interface between NbN/Pd and BLG, once NbN/Pd becomes superconducting [19]. Remarkably at smaller bias $I_{dc} < I_{SN}$, a drop in $dV/dI$ is observed. This region is amplified in Fig.

S9b and shown for several different $V_g$'s. We attribute the coherence peak-like features at $I_{SN}$ to the onset of an induced gap $\Delta'$ in the BLG area underneath the NbN/Pd contact, and the conductance enhancement below $I_{SN}$ to the resulting Andreev reflection at the superconducting BLG/normal BLG interface. In Fig. 4 of the main text, we show that the temperature dependence of $I_{SN}$ and $I_{JJ}$ (see main text for the definition of $I_{JJ}$) follow each other very well. Here we show in Fig. S9c the $V_g$-dependence of $I_{SN}$ and $I_{JJ}$. Again, $I_{SN}$ tracks $I_{JJ}$ very well, especially for the hole side. On the electron side, $I_{JJ}$ decreases more rapidly than $I_{SN}$, likely due to the growth of an interface barrier between the gapped BLG and normal BLG regions. The appearance of a small differential resistance peak at zero bias in Fig. S9b at $V_g = +3$ V validates this hypothesis, similar to what's observed in Ref. [6] for a single NbSe$_2$/BLG junction. Similar d$V$/d$I$ measurement performed using the other NbN/Pd contact reveals a similar $I_{SN}$, suggesting that both contacts develop a similar induced gap $\Delta'$.

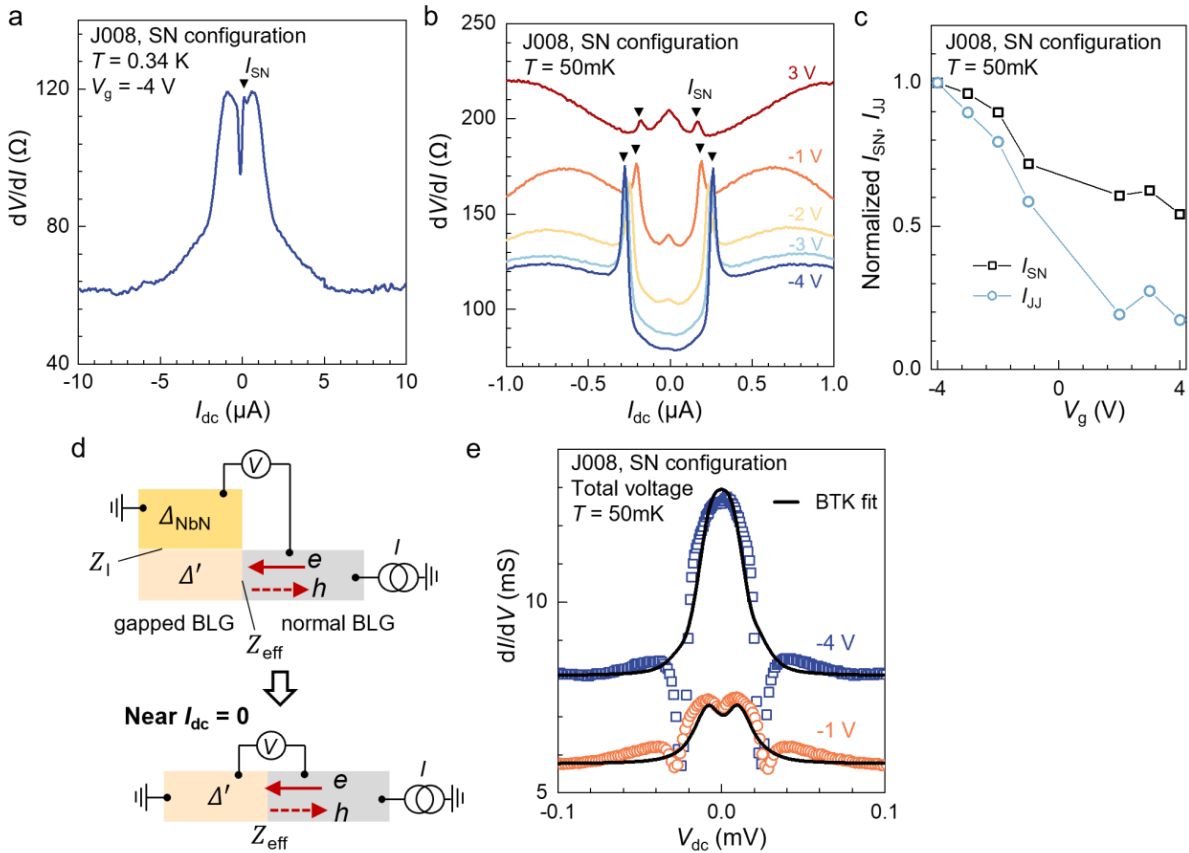

**Fig. S9 | Andreev reflections in a NbN/Pd-BLG SN junction. a,** Large-range $I_{dc}$ sweep of d$V$/d$I$ on a single SN junction in J008. **b,** d$V$/d$I$ versus $I_{dc}$ near the zero current bias at selected $V_g$'s. **c,** Normalized $I_{SN}$ and $I_{JJ}$ as a function of $V_g$. $T = 50$ mK. $I_{SN}$ and $I_{JJ}$ are normalized to their respective values at $V_g = -4$ V. **d,** Schematics of the two-gap model and the Andreev reflections at the two SN junctions. **e,** d$I$/d$V$ versus $V_{dc}$ near the zero bias at selected $V_g$'s, obtained by integrating the data in **b** to get $V(I)$ and then taking the first derivative. The black solid lines plot the corresponding BTK fits.

Figure S9d illustrates an effective two-gap model of our junction [6, 20], which consists of two SN junctions in series. We estimate the barrier strength of the SC-graphene interface $Z_I \sim 1$ from the conductance suppression in Fig. S9(a). This value limits the size of $\Delta'$ in the gapped BLG region. To obtain $\Delta'$, we convert the $dV/dI(I_{dc})$ data in Fig. S9b to differential conductance $dI/dV(V_{dc})$ [21-23], and fit it to the Blonder-Tinkham-Klapwijk (BTK) model [19]. Details of the fitting procedure can be found in our earlier work [24]. A few examples are shown in Fig. S9e. The fitting results of $\Delta'$ and the effective barrier strength $Z_{eff}$, as summarized in Fig. 4d of the main text.

## 10. Effect of vortex trapping in the NbN/Pd leads

In this section, we discuss the unusual effect of magnetic field and current on the supercurrent of the JJ. In Fig. S10a, the superconducting state of J008 completely vanishes after sweeping an external magnetic field. Similarly, after sweeping a large dc current through the JJ, the critical current becomes suppressed (See Fig. S10b). We note that the only way to restore the supercurrent is to thermal cycle the junction above the $T_c$ of NbN. These behaviors imply the trapping of vortices which are likely introduced by either an external field or the self-field of a current flow [25], serving as an indication of a disordered superconducting film. Improving the quality of NbN is necessary to fabricate high-performance and transparent Josephson junctions.

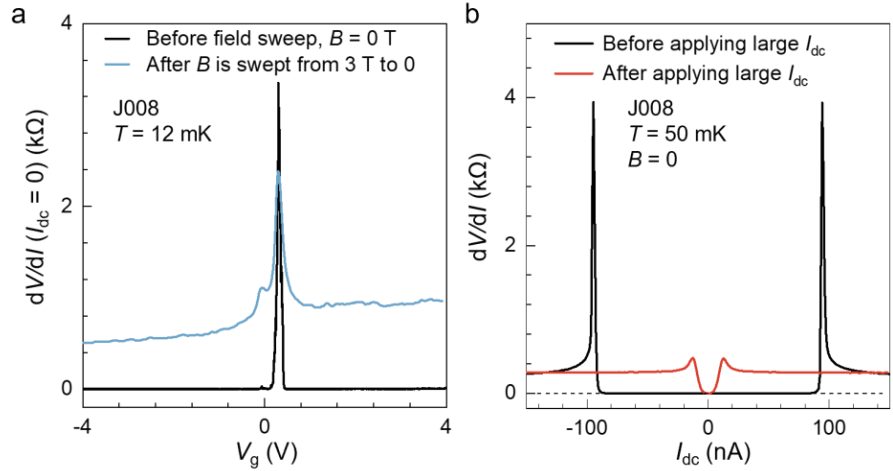

**Fig. S10 | Signature of magnetic field and current-induced vortex trapping. a**, Differential resistance $dV/dI$ at $I_{dc} = 0$ in J008 as a function of $V_g$ measured before and after the field sweeping. The field is swept from 3 T to zero at a rate of 1 mT/s, and the zero-resistance of the JJ vanishes afterwards. **b**, $dV/dI$ versus $I_{dc}$ measured before and after a large current bias sweep. The critical current of the JJ becomes suppressed after a dc bias current is swept from 10 µA to zero at a rate of 100 nA/s. The $dV/dI$ curves are taken as sweeping $I_{dc}$ within $\pm 200$ nA to eliminate Joule heating.

## 11. Making NbN/Pd via contacts to Bi$_2$Se$_3$

We have transferred NbN/Pd via contacts to Bi$_2$Se$_3$, an air-sensitive topological insulator inside a glovebox. Figure S11 shows the optical and AFM images of an exemplary stack. The transfer results in smooth and bubble-free stacks and demonstrates the general applicability of the via method to many air-sensitive and damage-sensitive materials, which expands the reach of the SC proximity effect.

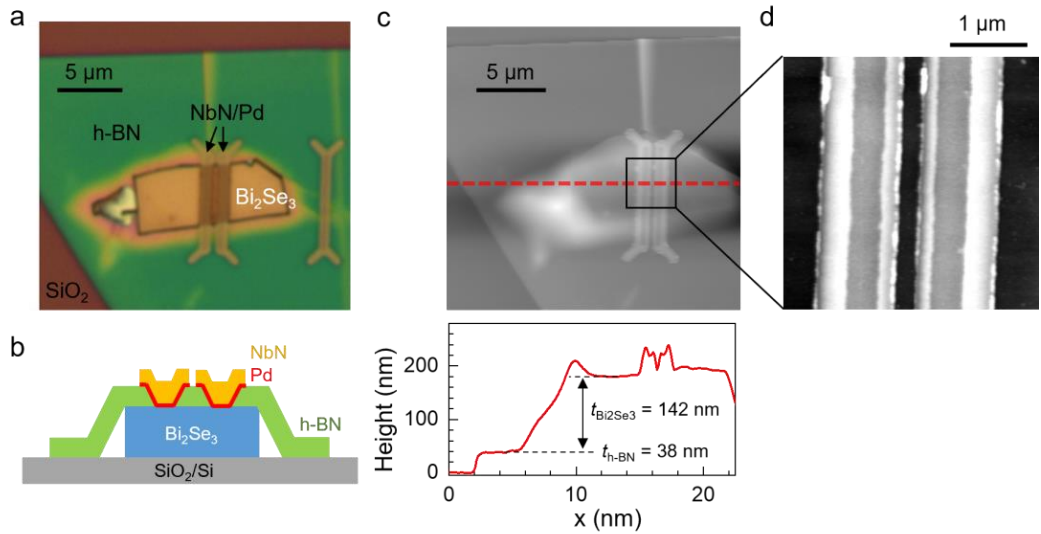

**Fig. S11 | NbN/Pd via contacts to air-sensitive Bi$_2$Se$_3$. a,** Optical image and **b,** schematic of a NbN/Pd-Bi$_2$Se$_3$ junction device. **c,** AFM image of the same device shown in **a**. The bottom panel plots the height profile along the red dashed line. The thickness of the Bi$_2$Se$_3$ flake is ~140 nm. **d,** Zoomed-in AFM image on the junction area with a clear gap of ~250 nm between the two via contacts.